\documentclass[11pt]{article}

\usepackage[margin=1in]{geometry}
\usepackage[T1]{fontenc}
\usepackage{lmodern}
\usepackage{microtype}

\usepackage{amsmath,amssymb}
\usepackage{algorithm}
\usepackage{algpseudocode}

\usepackage{graphicx}
\usepackage{booktabs}
\usepackage{subcaption}
\usepackage{threeparttable}

\usepackage[accsupp]{axessibility}

\usepackage[square,numbers,sort&compress]{natbib}

\usepackage{hyperref}
\hypersetup{hidelinks}

\newcommand{\degrees}{^{\circ}}
\newcommand{\proj}{\operatorname{proj}}
\newcommand{\dist}{\operatorname{dist}}
\newcommand{\angledeg}{\operatorname{angle}}

\title{Maritime Vessel Tracking}
\author{
  \begin{tabular}{c}
    John Mahlon Scott and
    Hsin-Hsiung Huang\thanks{Corresponding author: \texttt{hsin.huang@ucf.edu}} \\
    \\
    School of Data, Mathematical, and Statistical Sciences \\
    University of Central Florida, Orlando, FL 32816, USA
  \end{tabular}
}

\date{} 

\begin{document}
\maketitle

\begin{abstract}
The Automatic Identification System (AIS) provides time stamped vessel positions and kinematic reports that enable maritime authorities to monitor traffic. We consider the problem of relabeling AIS trajectories when vessel identifiers are missing, focusing on a challenging nationwide setting in which tracks are heavily downsampled and span diverse operating environments across continental U.S. waters. We propose a hybrid pipeline that first applies a physics-based screening step to project active track endpoints forward in time and select a small set of plausible ancestors for each new observation. A supervised neural classifier then chooses among these candidates, or initiates a new track, using engineered space time and kinematic consistency features. On held out data, this approach improves posit accuracy relative to unsupervised baselines, demonstrating that combining simple motion models with learned disambiguation can scale vessel relabeling to heterogeneous, high volume AIS streams.
\end{abstract}


\section*{List of Abbreviations}
\begin{table}[htbp]
\centering
\begin{tabular}{@{}ll@{}}
\toprule
Abbrev. & Meaning / Definition \\
\midrule
AIS   & Automatic Identification System (maritime transponder network) \\
ATD   & Algorithms for Threat Detection challenge \cite{chengAlgorithmsThreatDetection2025} \\
CBTR  & Physics-aware trajectory reconstruction method of \citet{chenUnsupervisedVesselTrajectory2023} \\
CTRV  & Constant turn rate and velocity magnitude modeling \\
COG   & Course over ground (degrees) \\
CV    & Constant velocity modeling \\
Posit & A single AIS position report with time stamped position and kinematics \\
SOG   & Speed over ground (knots) \\
Track & Historical trajectory of a given vessel, list of posits for one vessel \\
UTM   & Universal Transverse Mercator projection (metric $x/y$ coordinates) \\
KF    & Kalman filter \\
NN    & Nearest-neighbor association \\
\bottomrule
\end{tabular}
\end{table}

\section{Introduction}
The Automatic Identification System (AIS) is used by maritime authorities worldwide to monitor vessel positions. Vessels typically report their locations every 30 seconds to 6 minutes, depending on type and status. We aim to relabel daily AIS data with the correct vessel identifiers when they are missing. This problem was introduced as part of the Algorithms for Threat Detection (ATD) challenge posed in Summer 2025 \cite{chengAlgorithmsThreatDetection2025}. Throughout, we use the term \emph{AIS posit} to mean a single time-stamped AIS position report (latitude/longitude plus kinematics).

Unlabeled trajectory relabeling is often approached by unsupervised clustering with kinematic heuristics. Our prior work introduced a physics-aware clustering method (called Clustering based Trajectory Reconstruction, or CBTR) that reconstructs trajectories from localized AIS data near Norfolk, Virginia \cite{chenUnsupervisedVesselTrajectory2023}. The present study addresses a harder task: relabeling heavily downsampled tracks across all continental U.S. waters. This broader scope introduces substantial heterogeneity (vessel types, waterway geometries, traffic density), which challenges purely physics-based methods. To leverage historical data at scale, we recast the task as supervised classification, retaining CBTR's physical intuition while adding a neural classifier evaluated by \emph{posit accuracy} (Sec.~\ref{sec:experiments}; see also \cite{chengAlgorithmsThreatDetection2025}).

\begin{figure}[htbp]
    \centering
    \includegraphics[width=0.99\linewidth,
      alt={A map showing the coastline of Louisiana with 20 distinct vessel tracks overlaid. Each track is a series of points colored uniquely to represent a single vessel. Arrows indicate the direction of travel.}]{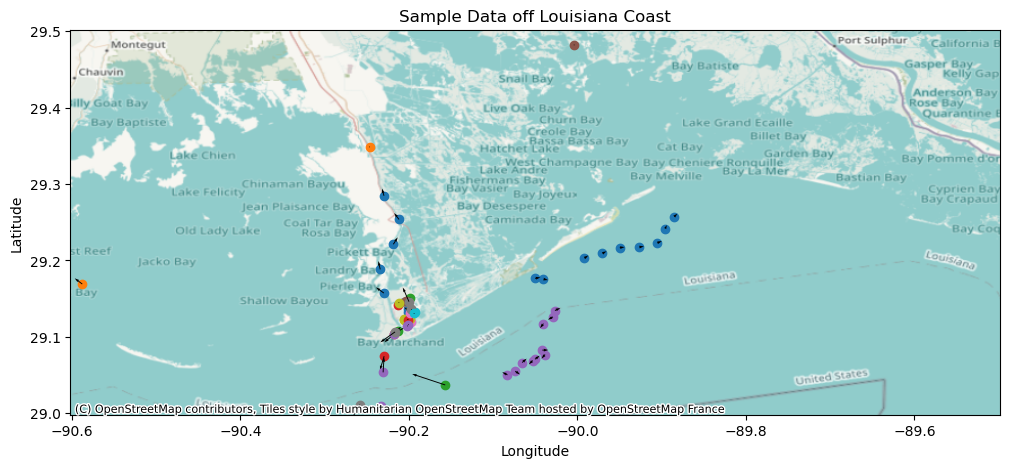}
    \caption{Twenty vessel tracks near the Louisiana coastline, colored by vessel identity. Arrows indicate direction of travel. The task is to restore correct colors (identities) when labels are absent.}
    \label{fig:sampdata}
\end{figure}

\section{Dataset and Metric}\label{sec:data}
Raw U.S. AIS data are publicly available from Marine Cadastre \cite{VesselTrafficMarine}. The portion used here spans 2{,}482 days and includes the following fields: a point index (\texttt{point\_id}); the Maritime Mobile Service Identity (MMSI), stored as \texttt{track\_id}; timestamp; latitude; longitude; course over ground (COG, degrees); and speed over ground (SOG, knots), as shown in Table~\ref{tab:datahead}. Compared to the localized Norfolk, VA, data analyzed in \citet{chenUnsupervisedVesselTrajectory2023}, these nationwide data present a markedly more challenging and realistic scenario. We downsample to roughly one report per vessel every 30 minutes, add slight spatiotemporal noise, and remove most stopped-vessel reports to focus on active paths. This preprocessed dataset is made available in \cite{Scott2025}.

\begin{table}[htbp]
\centering
\caption{Example data columns.}
\label{tab:datahead}
\begin{tabular}{@{}ccccccc@{}}
\toprule
point id & track id & time & lat & lon & speed (kn) & course ($\degrees$) \\
\midrule
0 & 338214987 & 00:00:00 & 28.033870 & -96.974543 & 6.5 & 56.1 \\
1 & 366970770 & 00:00:00 & 30.010361 & -90.726282 & 2.9 & 270.3 \\
2 & 368013670 & 00:00:01 & 30.708593 & -88.039977 & 0.0 & 156.5 \\
3 & 538002778 & 00:00:02 & 30.006844 & -90.472937 & 12.9 & 158.0 \\
4 & 368014410 & 00:00:03 & 28.174239 & -82.818806 & 2.5 & 74.8 \\
\bottomrule
\end{tabular}
\end{table}

\paragraph{Metric (posit accuracy).}
A \emph{posit} is a single AIS position report (time stamped position and kinematics). For each posit in a track, we award one point for correctly identifying its predecessor and one for correctly identifying its successor. The sum divided by the total available points yields the final \emph{posit accuracy}. See \cite{chengAlgorithmsThreatDetection2025} for details and examples.

\section{Mathematical Background}\label{sec:math}
We work in Universal Transverse Mercator (UTM) projected coordinates to simplify distance calculations. Let
\[
p_i \;=\; [\,t_i,\, x_i,\, y_i,\, v_i,\, \psi_i\,]
\]
denote time, UTM easting/northing $(x,y)$, speed $v$ (m/s), and course $\psi$ (radians, measured from the positive $x$ axis) for posit $i$. For UTM details, see \citet[p.~57]{snyderMapProjectionsWorking1987}. Write $\Delta t = t_j - t_i$.

\subsection{Coordinate conventions and local frame}\label{ssec:localframe}
We transform global errors into a local along-track and cross-track frame anchored at heading $\psi_i$. Let
\[
R(\psi_i) \;=\;
\begin{bmatrix}
\cos\psi_i & -\sin\psi_i \\
\sin\psi_i & \phantom{-}\cos\psi_i
\end{bmatrix},
\qquad
\text{so that}\quad
\begin{bmatrix}e_\parallel \\ e_\perp\end{bmatrix}
\;=\; R(\psi_i)^\top \!\begin{bmatrix}x_j-\widehat{x}_i\\ y_j-\widehat{y}_i\end{bmatrix}.
\]
Here $(\widehat{x}_i,\widehat{y}_i)$ is a kinematic projection of $p_i$ to time $t_j$ (Sec.~\ref{ssec:kinematics}). We also use the wrapped course difference
\[
\Delta_c \;=\; \mathrm{atan2}\!\big(\sin(\psi_j-\psi_i),\,\cos(\psi_j-\psi_i)\big)\in(-\pi,\pi],
\]
to avoid discontinuities at $2\pi$.

\subsection{Kinematic projection models}\label{ssec:kinematics}
We use continuous-time kinematics to project a state forward or backward. Under constant velocity (CV),
\begin{align}
x_{i\to j} &= x_i + v_i \cos\psi_i\, \Delta t, \\
y_{i\to j} &= y_i + v_i \sin\psi_i\, \Delta t, \\
v_{i\to j} &= v_i, \qquad \psi_{i\to j} = \psi_i.
\end{align}
Under constant turn rate $\omega$ with constant speed magnitude (CTRV),
\begin{align}
x_{i\to j} &= x_i + \frac{v_i}{\omega}\!\left[\sin(\psi_i + \omega \Delta t) - \sin\psi_i\right], \\
y_{i\to j} &= y_i + \frac{v_i}{\omega}\!\left[-\cos(\psi_i + \omega \Delta t) + \cos\psi_i\right], \\
v_{i\to j} &= v_i, \qquad \psi_{i\to j} = \psi_i + \omega \Delta t,
\end{align}
with the $\omega\to 0$ limit recovering CV. To model speed changes without turn, we allow a tangential constant acceleration $a$ along heading $\psi_i$:
\begin{align}
v_{i\to j} &= v_i + a\,\Delta t, \\
x_{i\to j} &= x_i + \Big(v_i\,\Delta t + \tfrac{1}{2}a\,\Delta t^2\Big)\cos\psi_i, \\
y_{i\to j} &= y_i + \Big(v_i\,\Delta t + \tfrac{1}{2}a\,\Delta t^2\Big)\sin\psi_i .
\end{align}
We denote the projection compactly by $\widehat{p}_i = \proj(p_i,\Delta t)$ and the backward projection by $\widehat{p}_j = \proj(p_j,-\Delta t)$.

\subsection{Error model, anisotropic covariance, and ellipsoidal gating}\label{ssec:maha}
Given a projection $\widehat{p}_i$, define the displacement error $e = [x_j-\widehat{x}_i,\; y_j-\widehat{y}_i]^\top$ and its local components $(e_\parallel,e_\perp)$ from Sec.~\ref{ssec:localframe}. We score kinematic consistency with an anisotropic Gaussian model
\[
\begin{bmatrix}e_\parallel \\ e_\perp\end{bmatrix}
\sim \mathcal{N}\!\Big(\begin{bmatrix}0\\0\end{bmatrix},\;
\Sigma(\Delta t)\Big), \qquad
\Sigma(\Delta t)=
\begin{bmatrix}
\sigma_\parallel^2(\Delta t) & 0\\[2pt]
0 & \sigma_\perp^2(\Delta t)
\end{bmatrix},
\]
where, in practice, $\sigma_\parallel^2(\Delta t)=\sigma_0^2+\alpha_\parallel^2\Delta t^2$ and $\sigma_\perp^2(\Delta t)=\sigma_0^2+\alpha_\perp^2\Delta t^2$ with $\alpha_\parallel > \alpha_\perp$ (larger uncertainty along track). The induced Mahalanobis distance
\[
M^2 \;=\; e^\top \big(R(\psi_i)\,\Sigma^{-1}R(\psi_i)^\top\big)\,e
\]
yields the ellipsoidal gate
\begin{equation}
\label{eq:gate-ellipse}
M^2 \;\le\; \tau^2,
\end{equation}
with threshold $\tau$ tuned to trade recall versus precision in candidate screening.

\subsection{Heading and turn-rate consistency (angular gating)}\label{ssec:anglegate}
Let the observed step be $\Delta_{\text{actual}} = [x_j-x_i,\; y_j-y_i]^\top$ and the predicted step be $\Delta_{\text{proj}} = [\widehat{x}_i-x_i,\; \widehat{y}_i-y_i]^\top$. We enforce a dynamic-feasibility gate on the turn between them:
\begin{equation}
\label{eq:gate-angle}
\angledeg(\Delta_{\text{actual}}, \Delta_{\text{proj}}) \;\le\; \theta,
\qquad \theta \in [60\degrees,\,90\degrees]\ \ (\text{we use } \theta\approx 85\degrees).
\end{equation}
We also use the implied speed needed to connect $i$ to $j$:
\[
v^{\text{req}} \;=\; \frac{\lVert \Delta_{\text{actual}} \rVert_2}{\Delta t}.
\]

\subsection{Screening as approximate MAP scoring}\label{ssec:map}
A fast approximate MAP score for the hypothesis ``$j$ follows $i$'' is
\begin{equation}
\label{eq:score}
s(i\!\to\!j) \;=\;
\underbrace{e^\top \big(R\Sigma^{-1}R^\top\big) e}_{\text{ellipsoidal displacement}}
\;+\; \lambda_\psi\,\Delta_c^2
\;+\; \lambda_v\,\big(v^{\text{req}}-v_j\big)^2
\;-\; \log \pi_{\text{cont}}(i),
\end{equation}
subject to the gates \eqref{eq:gate-ellipse} and \eqref{eq:gate-angle}. The ``New Vessel'' hypothesis uses
\begin{equation}
\label{eq:new}
s_{\text{new}}(j) \;=\; -\log \pi_{\text{birth}}(j),
\end{equation}
with priors $\pi_{\text{cont}}$ and $\pi_{\text{birth}}$ depending on local endpoint density and time gaps. The screening stage returns the top-$k$ hypotheses among $\{i\!\to\!j\}$ and New Vessel, which are then passed to the supervised classifier (Sec.~\ref{sec:method}).

\paragraph{Complexity.}
If $E_t$ endpoints are active at time $t$, screening evaluates $O(E_t)$ scores of the closed form \eqref{eq:score} and keeps the best $k$. Classifier cost is $O(k)$ per posit.

\section{Unsupervised Methods}\label{sec:unsup}
Classical methods for labeling data of this sort rely on kinematic consistency: given two posits $p_i$ and $p_j$ with $t_j>t_i$, we project $p_i$ forward to $t_j$ (and or $p_j$ backward to $t_i$), measure the mismatch, and prefer links with small errors that pass simple dynamical gates.

For algorithmic simplicity, we assume standard matrix-vector broadcasting rules. For example, whenever $P - p_i$ appears with $P$ an $n\times d$ matrix (stack of $n$ posits with $d$ fields) and $p_i$ a $1\times d$ row, $p_i$ is first repeated $n$ times along the first axis before subtraction.

\subsection{CBTR (physics-aware greedy linking)}
\begin{algorithm}[htbp]
\caption{CBTR distance computation and greedy linking}\label{alg:cbtr}
\begin{algorithmic}[1]
    \State \textbf{Input:} rows $P = [p_0, p_1, \dots, p_{n-1}]$ sorted by time
    \State Initialize distance matrix $D \gets \infty$ of size $n\times n$
    \For{$i = 0$ \textbf{to} $n-2$}
        \For{$j = i+1$ \textbf{to} $n-1$}
            \State $\Delta t \gets t_j - t_i$
            \State $\widehat{p}_i \gets \proj(p_i, \Delta t)$ \Comment{forward project $p_i$}
            \State $\widehat{p}_j \gets \proj(p_j, -\Delta t)$ \Comment{backward project $p_j$}
            \State $\Delta_{\text{actual}} \gets [x_j - x_i,\, y_j - y_i]$
            \State $\Delta_{\text{proj}} \gets [\widehat{x}_i - x_i,\, \widehat{y}_i - y_i]$
            \If{$\angledeg(\Delta_{\text{actual}}, \Delta_{\text{proj}}) \le \theta$}
                \State $d_f \gets \dist(\widehat{p}_i, p_j)$;\; $d_b \gets \dist(\widehat{p}_j, p_i)$
                \State $D[i,j] \gets \frac{1}{2}\,(d_f + d_b)$
            \EndIf
        \EndFor
    \EndFor
    \State \textbf{Linking:} process points chronologically, assigning each point to the nearest compatible previous track endpoint according to $D$.
\end{algorithmic}
\end{algorithm}

In CBTR \cite{chenUnsupervisedVesselTrajectory2023}, for each pair $(p_i,p_j)$ we project $p_i$ forward over $\Delta t$ and $p_j$ backward over $-\Delta t$, compute both distances, and average them. We additionally constrain turn rate by requiring that the angle between the predicted and observed displacements not exceed $\theta$ (we use about $85\degrees$). In practice, one also limits candidate pairs to a time window (for example a few hours) to reduce runtime. After distances are computed, points are processed in chronological order and each new point links to the nearest compatible past endpoint.

As reported later, CBTR attains about $0.33$ posit accuracy on our hold-out data.

\subsection{ATD 2025 baseline model}
The ATD 2025 Challenge \cite{chengAlgorithmsThreatDetection2025} provided a baseline model for comparison (originally developed by RAND). Between observations, vessels accelerate from initial speed to a target $v^{*}$, then decelerate back down to the end velocity. With $t^{*}$ the time at which $v^{*}$ is reached, a fitted $m^{*}$ (acceleration) satisfies the traversal constraint. The distance metric is
\begin{equation}\label{eqn:basedist}
\begin{aligned}
    d(p_2,p_1) \;=\;
    &\left\lvert \left(m^* - \frac{v_2 - v_1}{\Delta t}\right)\,(t^* - t_1) \right\rvert
    \;+\; \frac{\Delta_c}{\Delta t} \\
    &\;+\; \left\lvert \left(\frac{\Delta_{c-}}{\Delta_{t-}} - \frac{\Delta_c}{\Delta t}\right)\,\frac{1}{\Delta_{t-}} \right\rvert,
\end{aligned}
\end{equation}
where $\Delta t = t_2 - t_1$, $\Delta_c$ is the wrapped course difference from Sec.~\ref{ssec:localframe}, and $(\cdot)_{-}$ refers to the previous segment in the same track. This baseline achieves about $0.44$ posit accuracy.

\section{Our Methodology}\label{sec:method}
We adopt a hybrid approach inspired by CBTR. For each new query posit $q$ at time $t_q$, an unsupervised screening step projects active track endpoints forward and scores them using the anisotropic, angle-aware criteria from Sec.~\ref{sec:math}. The screen returns the top $k=16$ plausible ancestors together with the New Vessel hypothesis. A supervised classifier then makes the final assignment among these $k+1$ options.

\subsection{Physics-based candidate screening}
For each active endpoint $c \in \texttt{endpts}$ we compute displacement error $e$, wrapped course difference $\Delta_c$, and implied speed $v^{\text{req}}$. We apply the gates $M^2\le\tau^2$ and $\angledeg(\cdot,\cdot)\le\theta$ (Eqs.~\eqref{eq:gate-ellipse} and \eqref{eq:gate-angle}), then rank by the approximate MAP score from Eq.~\eqref{eq:score}. We keep the $k$ best candidates and always include New Vessel with score $s_{\text{new}}(q)$ from Eq.~\eqref{eq:new}.

\subsection{Supervised assignment model}
For each candidate $c_\ell$ we build a feature vector $\phi(q,c_\ell)\in\mathbb{R}^{231}$ comprising:
(i) local-frame residuals $(e_\parallel,e_\perp)$, time gap $\Delta t$, implied $v^{\text{req}}$ and turn rate;
(ii) forward and backward projection errors and their symmetrized averages;
(iii) heading and speed consistency and jitter indicators; and
(iv) context features such as endpoint density and time-of-day.
A fully connected network with three hidden layers (widths $2000$, $2000$, and $1000$) produces logits for the $k$ candidates and the New Vessel class. We train with cross-entropy; at inference we apply temperature scaling for calibrated probabilities.

\begin{algorithm}[htbp]
\caption{Candidate screening + supervised assignment}\label{alg:track}
\begin{algorithmic}[1]
    \State $endpts \gets [\,]$ \Comment{active track endpoints}
    \State $t \gets 0$ \Comment{next new track ID}
    \For{$i = 0$ \textbf{to} $n-1$} \Comment{$p_i$ processed in increasing time}
       \State $C \gets \text{screen}(p_i, endpts, k{=}16)$ \Comment{physics-based top-$k$ ancestors}
       \State $res \gets \text{classify}\big(p_i, C \cup \{\text{New Vessel}\}\big)$
       \If{$res = \text{New Vessel}$}
          \State $endpts[t] \gets p_i$;\; $\text{label}(p_i, t)$;\; $t \gets t+1$
       \Else
          \State $j \gets \text{index of ancestor in } endpts \text{ corresponding to } res$
          \State $endpts[j] \gets p_i$;\; $\text{label}(p_i, j)$
       \EndIf
    \EndFor
\end{algorithmic}
\end{algorithm}

\subsection{Oracle upper bound}
\begin{figure}[htbp]
    \centering
    \includegraphics[width=0.99\linewidth,
      alt={Map with vessel tracks colored by the oracle model's posit accuracy: yellow perfect, blue partial, purple zero.}]{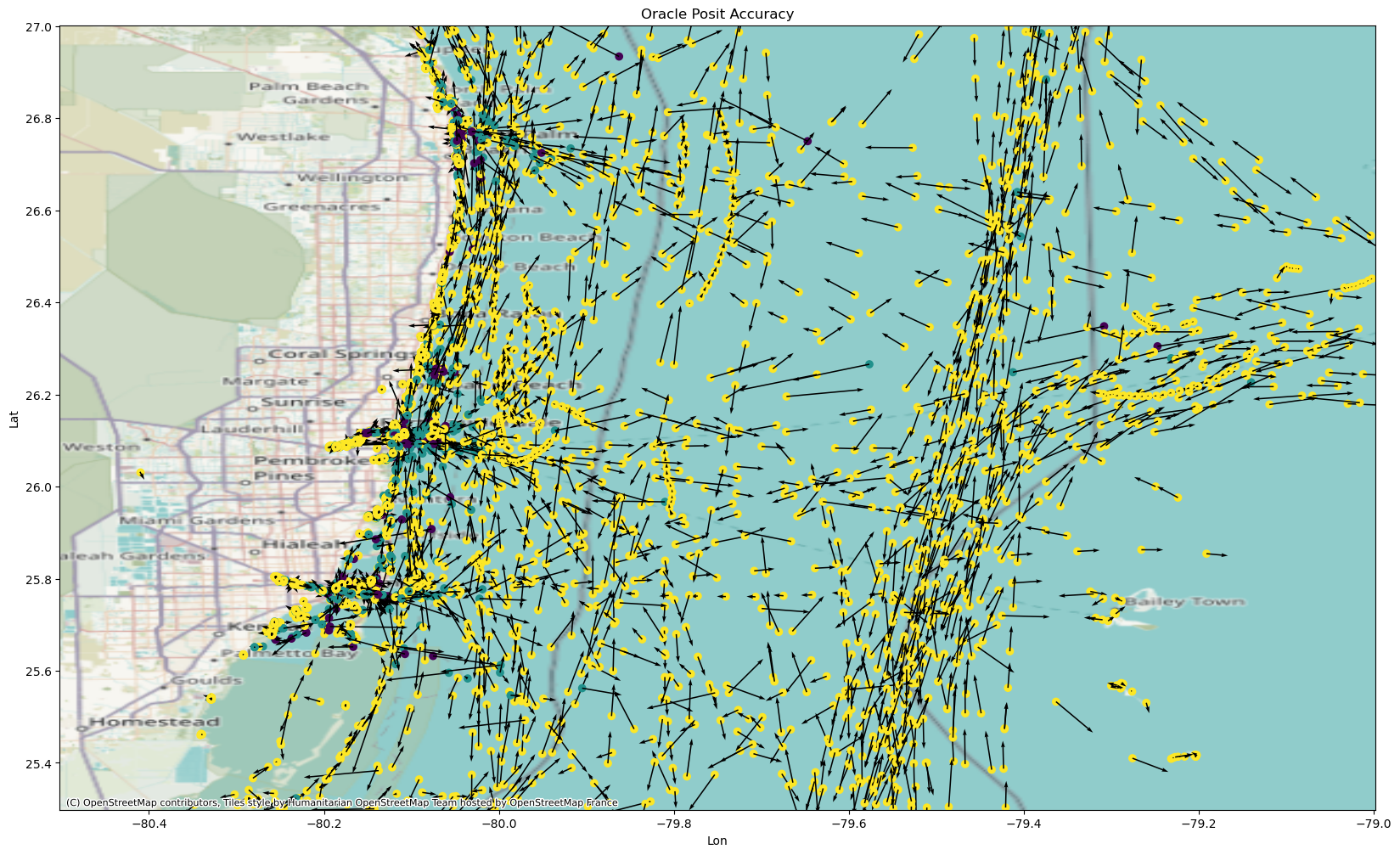}
    \caption{Oracle posit accuracy. Yellow indicates both neighbors correct; blue indicates one neighbor correct; dark purple indicates neither correct.}
    \label{fig:oposit}
\end{figure}

An oracle that always selects the true class among the $k+1$ screened options attains an empirical ceiling near $0.85$ posit accuracy for $k=16$ on our data (Fig.~\ref{fig:oposit}). This gap indicates headroom for improved learned disambiguation and or smarter screening in dense ports.

\section{Experiments}\label{sec:experiments}
\subsection{Training details and throughput}
Computation for this model is inherently sequential in the time dimension. To make data generation feasible, we parallelize feature generation across a 1{,}600-core compute cluster to compute the full classification dataset for all 2{,}482 days in under an hour. The classifier trains on an H100 GPU for about 120 epochs (about 2 days) using Adam with large effective batches (about 1000), mixed precision, gradient clipping ($\ell_2 \le 1$), and label smoothing ($\epsilon=0.05$). As discussed by \citet{kandelEffectBatchSize2020,wilsonMarginalValueAdaptive2017}, smaller batches and SGD-like optimizers may generalize better. In our setting we found that reducing Adam's learning rate as data and model scales grew was necessary for stability. We calibrate post hoc with temperature scaling on a held-out validation set.

\subsection{Baselines}
In addition to CBTR and the ATD 2025 baseline, we evaluated standard multi-target tracking and clustering baselines under the same preprocessing and scoring:
KF(CV) + NN; KF(CTRV) + NN; adaptive KF(CV) + NN; DBSCAN; and linkage clustering on the CBTR distance matrix. Kalman filter process and measurement noise and gating thresholds are tuned on validation days only.

\subsection{Scenario coverage}
We stratify evaluation by geography and density:
open water (offshore); coastal approaches and river channels; and ports and docks. Accuracy is highest offshore and degrades in congested port and river systems, with errors concentrated at docks.

\begin{figure}[htbp]
    \centering
    \includegraphics[width=0.99\linewidth,
      alt={Map with tracks colored by the trained classifier's posit accuracy: high accuracy offshore; lower accuracy in congested port regions.}]{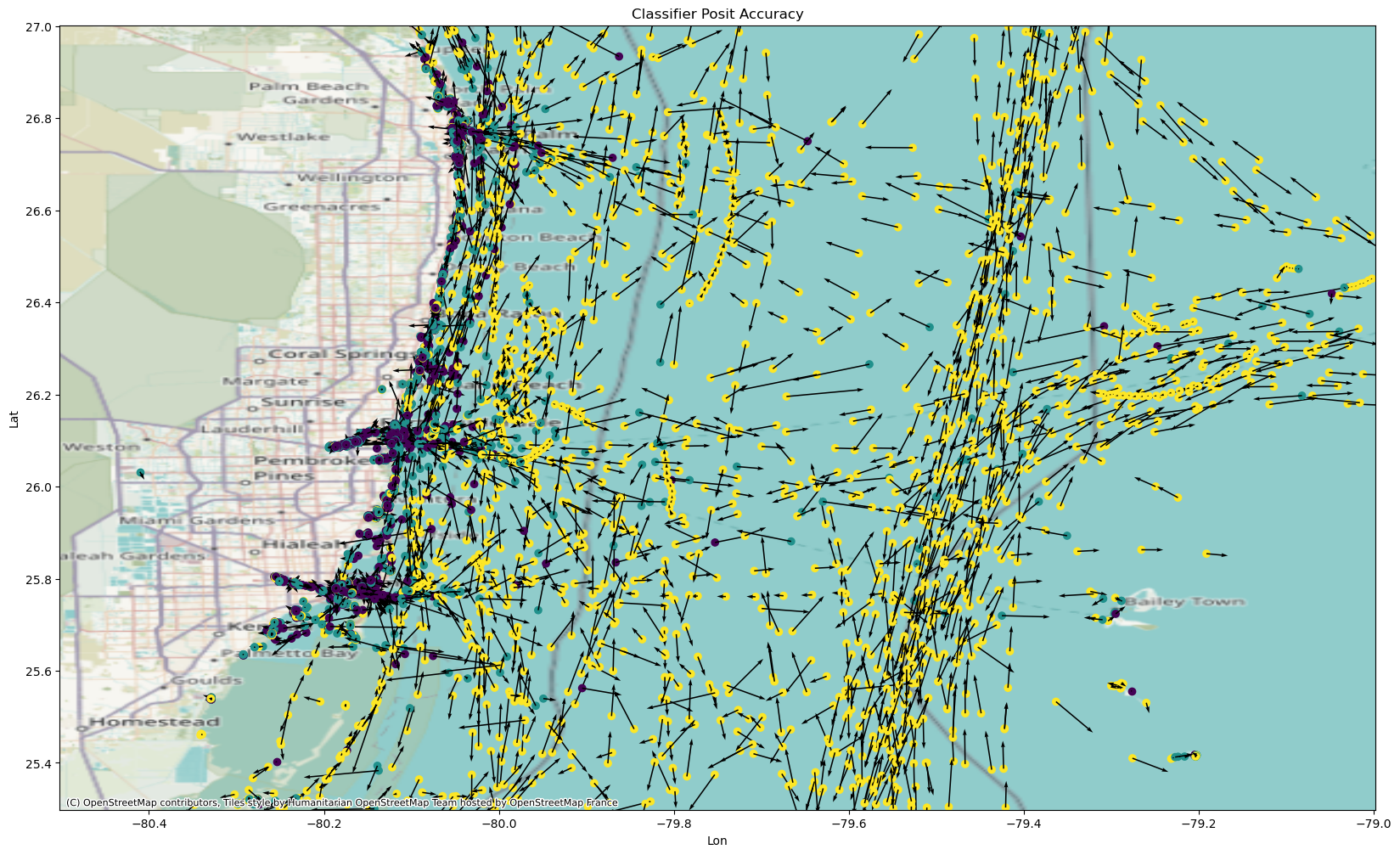}
    \caption{Classifier posit accuracy. Accuracy is high offshore and degrades in congested port and river systems.}
    \label{fig:cposit}
\end{figure}

\begin{figure}[htbp]
\centering
\begin{subfigure}[b]{0.48\linewidth}
  \centering
  \includegraphics[width=\linewidth,
    alt={Map with tracks colored by the trained classifier's posit accuracy: low accuracy in Miami port area.}]{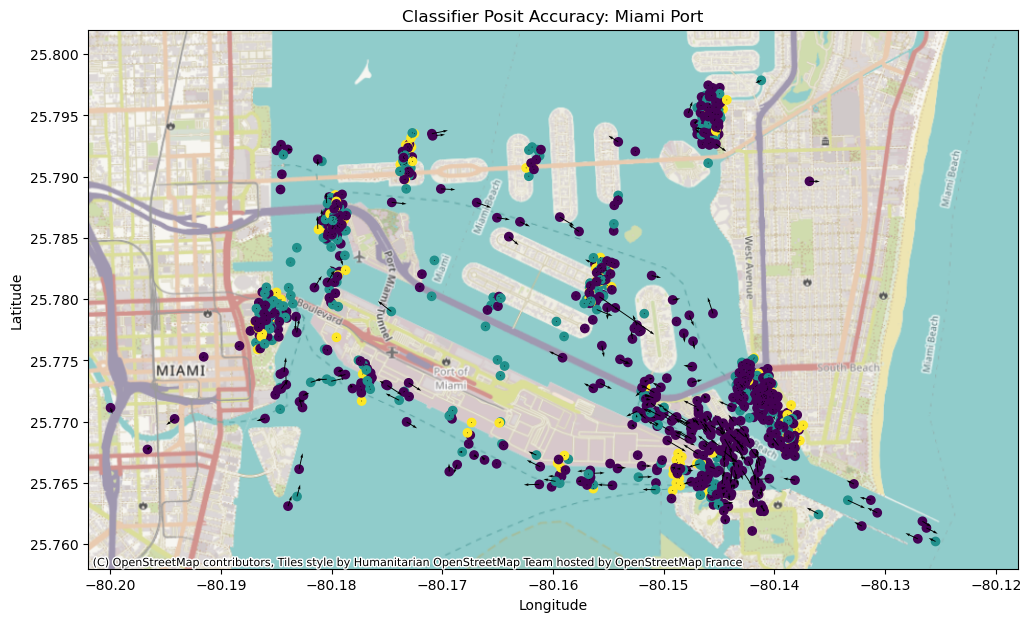}
  \caption{Classifier posit accuracy in the Port of Miami region}
  \label{fig:miamiPosit}
\end{subfigure}\hfill
\begin{subfigure}[b]{0.48\linewidth}
  \centering
  \includegraphics[width=\linewidth,
    alt={Satellite imagery of Miami Port area with vessel docks highlighted.}]{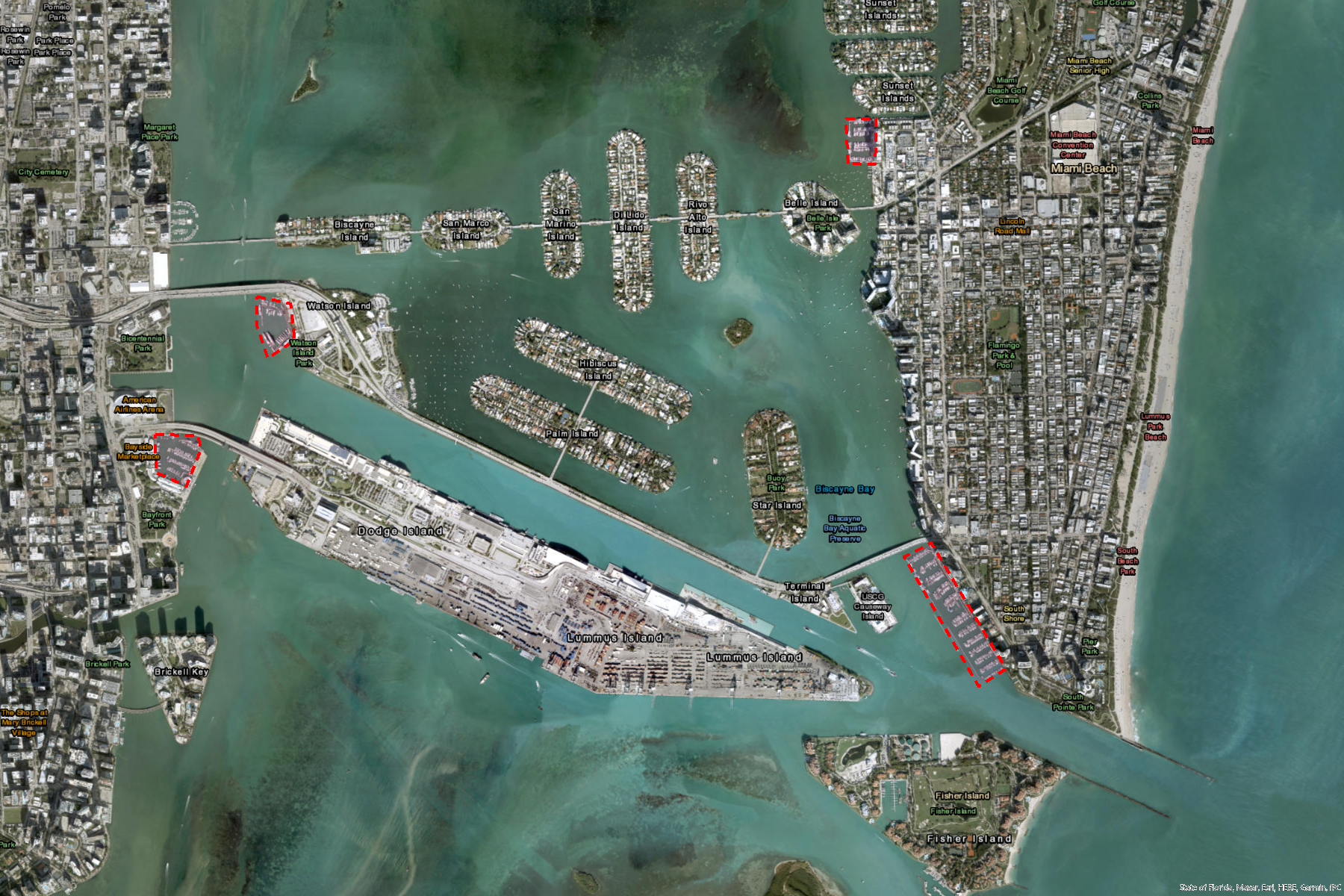}
  \caption{Satellite imagery of the Port of Miami with docks highlighted}
  \label{fig:miamiPort}
\end{subfigure}
\caption{Error concentration in a high density port zone (Port of Miami).}
\end{figure}

\subsection{Main results}
\begin{figure}[htbp]
    \centering
    \includegraphics[width=0.7\linewidth,
      alt={Plot showing simulated posit accuracy as a function of simulated classifier accuracy, with the proposed model marked.}]{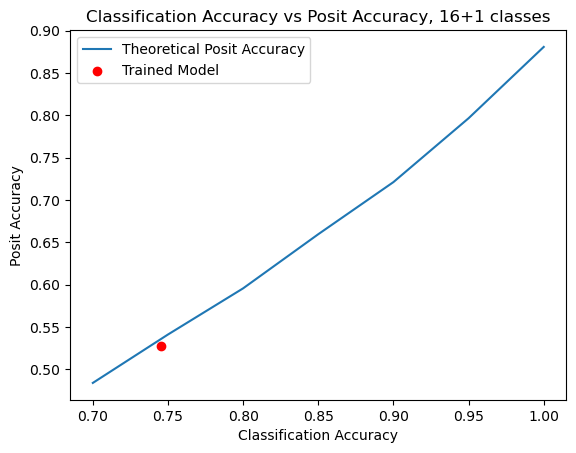}
    \caption{Posit accuracy for various simulated classifier accuracies. Our model is shown for context.}
    \label{fig:classPositAcc}
\end{figure}

Trained as above, the classifier reaches mean accuracy and F1 of about $75\%$ on held-out classification, translating to overall posit accuracy of $0.53$. The relationship between posit accuracy and classification accuracy is shown in Fig.~\ref{fig:classPositAcc}.

\begin{table}[htbp]
\centering
\caption{Test-set classification report.}
\label{tab:classreport}
\begin{threeparttable}
\small
\begin{tabular}{@{}lccc@{}}
\toprule
Class & Precision & Recall & F1 \\
\midrule
1   & 0.94 & 0.84 & 0.89 \\
2   & 0.90 & 0.85 & 0.87 \\
3   & 0.87 & 0.87 & 0.87 \\
4   & 0.84 & 0.86 & 0.85 \\
5   & 0.85 & 0.83 & 0.84 \\
6   & 0.87 & 0.78 & 0.82 \\
7   & 0.89 & 0.73 & 0.80 \\
8   & 0.90 & 0.70 & 0.79 \\
9   & 0.88 & 0.68 & 0.77 \\
10  & 0.87 & 0.66 & 0.75 \\
11  & 0.79 & 0.63 & 0.70 \\
12  & 0.71 & 0.58 & 0.64 \\
13  & 0.65 & 0.55 & 0.59 \\
14  & 0.61 & 0.52 & 0.56 \\
15  & 0.51 & 0.55 & 0.53 \\
16  & 0.47 & 0.62 & 0.54 \\
New & 0.67 & 0.67 & 0.67 \\
\midrule
\textbf{Average} & \textbf{0.77} & \textbf{0.70} & \textbf{0.73} \\
\bottomrule
\end{tabular}
\begin{tablenotes}
\small
\item Overall classification accuracy is 0.7452.
\end{tablenotes}
\end{threeparttable}
\end{table}

\begin{table}[htbp]
\centering
\caption{Test-set posit accuracy.}
\label{tab:positAcc}
\begin{tabular}{@{}lc@{}}
\toprule
Model & Posit Accuracy \\
\midrule
CBTR     & 0.33 \\
ATD 2025 baseline & 0.44 \\
\textbf{Ours (hybrid)} & \textbf{0.53} \\
\bottomrule
\end{tabular}
\end{table}

\section{Discussion}
Accuracy is high in open waters where trajectories are well separated, but degrades in dense, multi-vessel environments (ports and rivers) with many plausible ancestors at small spatiotemporal separations. In ports, spatiotemporal noise and downsampling can make a vessel appear to shift slips, causing confusion with a neighboring docked vessel. In channels, accuracy also drops under high density. To materially improve posit accuracy, a focused effort on high density regions is needed. Incorporating waterway geometry (navigable polygons, fairways) and port-aware priors is a promising next step.

The oracle analysis helps separate two bottlenecks: screening may exclude the true ancestor in dense zones, and classification may fail to disambiguate collisions among plausible candidates. Closing the gap to the oracle ceiling likely requires both improved screening (for example, geometry-aware continuation priors) and richer learned disambiguation (for example, context features tied to port structure and stop and go behavior).

\section{Conclusion}
A hybrid supervised approach that preserves CBTR's physics-based screening achieves scalable relabeling on nationwide AIS data. While the $0.53$ posit accuracy is modest, it reflects a substantially harder setting than the localized Norfolk study where unsupervised CBTR was near-perfect \cite{chenUnsupervisedVesselTrajectory2023}. Future work should focus on improving accuracy in high density regions (ports and docks) and incorporating waterway constraints and vessel behavior priors to improve performance in high traffic environments.

\paragraph{Reproducibility and code.}
All preprocessing, screening, and training scripts are publicly released at
\href{https://github.com/KingJMS1/ucf_atd2025}{https://github.com/KingJMS1/ucf\_atd2025}.
The ATD 2025 baseline is released separately at
\href{https://gitlab.com/algorithms-for-threat-detection/2025/atd2025}{https://gitlab.com/algorithms-for-threat-detection/2025/atd2025}.
The dataset used to train the neural model is released in \cite{Scott2025}.

\bibliographystyle{plainnat}
\bibliography{AIS2025}

\end{document}